\documentclass[amsmath,floatfix,twocolumn,superscriptaddress,citeautoscript]{revtex4-1}
\usepackage{subfigure}
\usepackage{siunitx}
\usepackage{amssymb}
\usepackage{amsmath}
\usepackage{graphicx}
\usepackage{array}
\usepackage{dcolumn}
\usepackage{psfrag}
\usepackage{bm}
\usepackage{color}
\usepackage{multirow}

\setcitestyle{super}

\begin{document}

\title{Light-facilitated ferroelectric switching in wurtzite crystals}

\author{Ricardo Jim\'enez-S\'anchez}
\affiliation{Instituto Polit\'ecnico Nacional, ESIME-Culhuac\'an, Av. Santa Ana 1000, Ciudad de M\'exico C.P. 04440, Mexico}
\affiliation{Institut de Ciència de Materials de Barcelona (ICMAB-CSIC), Carrer dels Til·lers, 08193 Cerdanyola del Vallès, 
Spain}

\author{Fernando Salazar}
\affiliation{Instituto Polit\'ecnico Nacional, ESIME-Culhuac\'an, Av. Santa Ana 1000, Ciudad de M\'exico C.P. 04440, Mexico}

\author{Miguel Cruz-Irisson}
\affiliation{Instituto Polit\'ecnico Nacional, ESIME-Culhuac\'an, Av. Santa Ana 1000, Ciudad de M\'exico C.P. 04440, Mexico}

\author{Riccardo Rurali}
\email{rrurali@icmab.es}
\affiliation{Institut de Ciència de Materials de Barcelona (ICMAB-CSIC), Carrer dels Til·lers, 08193 Cerdanyola del Vallès, 
Spain}

\author{Claudio Cazorla}
\email{claudio.cazorla@upc.edu}
    \affiliation{Departament de Física, Universitat Politècnica de Catalunya, 08034 Barcelona, Spain}
    \affiliation{Research Center in Multiscale Science and Engineering, Universitat Politècnica de Catalunya,
                 Campus Diagonal-Besòs, Av. Eduard Maristany 10–14, 08019 Barcelona, Spain}
    \affiliation{Institució Catalana de Recerca i Estudis Avançats (ICREA), Passeig Lluís Companys 23, 08010 Barcelona, Spain}

\begin{abstract}
Wurtzite ferroelectrics combine large remanent polarization with full CMOS compatibility, positioning them as a leading platform 
for next-generation non-volatile memory. Their practical deployment, however, is hindered by an intrinsically large coercive field, 
rooted in the high energy barrier separating the polar wurtzite phase from the nonpolar hexagonal phase that mediates polarization 
switching. Here, using first-principles calculations, we propose an alternative, field-free strategy for lowering this barrier: 
above-bandgap electronic photoexcitation. Taking LaN as a representative wurtzite ferroelectric, we show that light-induced carriers 
dramatically reduce the energy difference between the hexagonal intermediate phase and the wurtzite ground state, sharply reducing 
the energy barrier to ferroelectric switching. This effect originates from a photoinduced partial metallization of the polar phase, 
which screens the dipole-dipole interactions that stabilize ferroelectric order and thereby favors the competing nonpolar structure. 
The robustness of this mechanism is further confirmed for the rocksalt polymorph. Our results establish light as a powerful, non-invasive 
route to controlling ferroelectric switching in wurtzites, opening a path toward faster, lower-voltage, and more energy-efficient 
non-volatile memory technologies.
\end{abstract}

\maketitle

Wurtzites (WZ, space group $P6_3mc$) have rapidly evolved from a purely piezoelectric materials family into a genuinely new class 
of ferroelectrics. Archetypal WZ ferroelectrics, aluminum scandium nitride alloys (Al$_x$Sc$_{1-x}$N, ASN), exhibit switchable 
polarization of $0.8$--$1.7$~C$\cdot$m$^{-2}$ and retain their ferroelectric performance up to temperatures of $\sim 1{,}100$~$^\circ$C 
\cite{intro1,intro2}. Crucially, wurtzites combine this large remanent polarization with full compatibility with complementary 
metal-oxide-semiconductor (CMOS) technology and back-end-of-line integration \cite{intro3}, positioning them as prime candidates 
for next-generation non-volatile memories and neuromorphic devices.

\begin{figure*}
    \centering
        \includegraphics[width=0.9\textwidth]{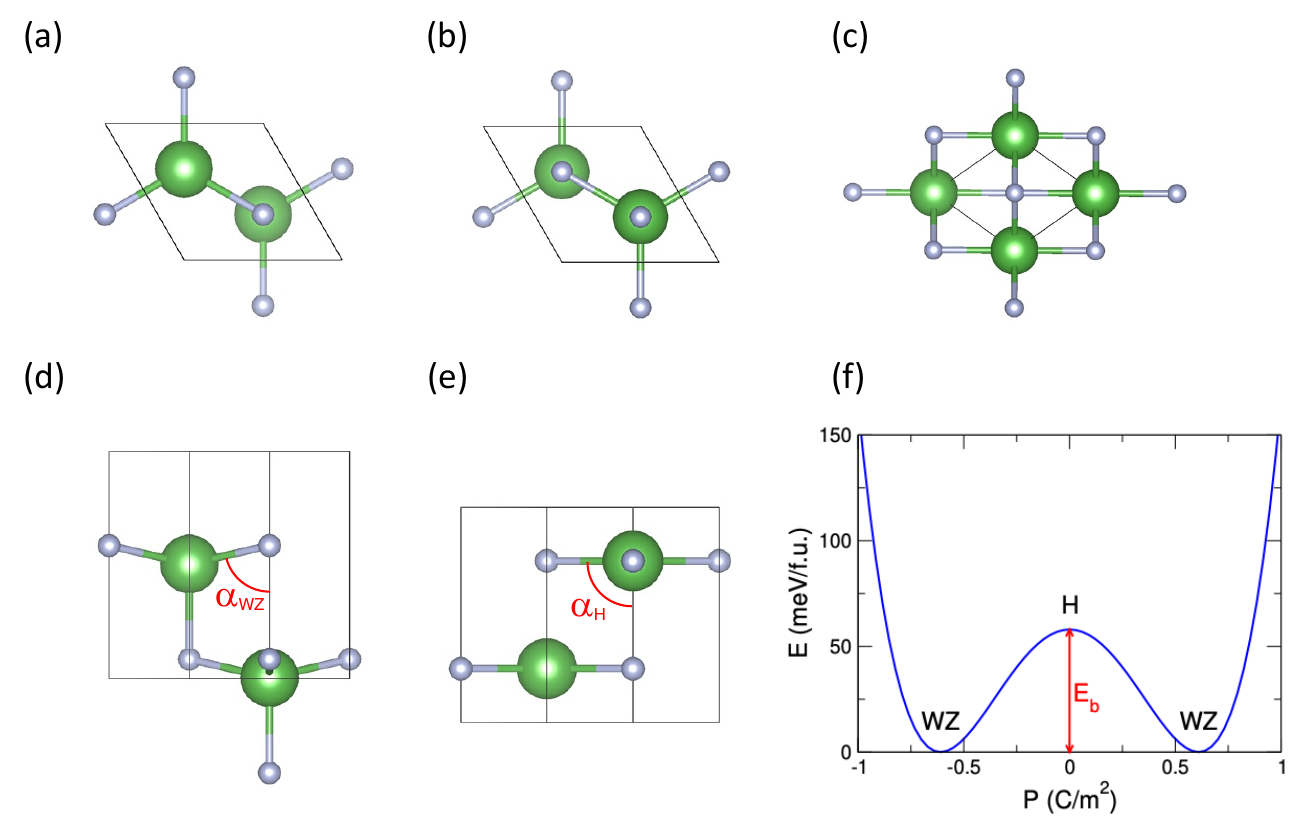}
        \caption{(a)--(d)~The wurtzite (WZ) ground-state, (b)--(e)~the layered hexagonal (H) saddle point of field-induced ferroelectric 
         switching in WZ, and (c)~the rocksalt (RS) structure. Large green and small gray spheres represent La and N atoms, respectively, 
         and solid black lines represent primitive cells. The angle $\alpha$ can be used to track the transformation of the WZ into the 
         H phase under illumination ($\alpha_{\rm WZ}=78.7^{\circ}$ and $\alpha_{\rm H}=90^{\circ}$). (f)~Double-well potential curve in 
         which the H phase appears as a saddle point separating two WZ structures with opposed polarization.}
        \label{fig1}
\end{figure*}

Over the past few years, significant progress has been made in overcoming the endurance and leakage bottlenecks that initially 
limited WZ-based devices. Whereas early ASN capacitors were restricted to roughly $10^{7}$ switching cycles by coercive fields 
exceeding $3$~MV~cm$^{-1}$, controlled partial-polarization switching has recently pushed the write endurance beyond $10^{10}$ 
cycles \cite{intro4}. In parallel, ASN ferroelectric diode memories have been demonstrated at diameters below $50$~nm \cite{intro5}, 
and alloy engineering is now being exploited to further tune the coercive field and phase stability \cite{intro6}. Together, these 
advances mark the transition of WZ ferroelectrics from a scientific curiosity into a viable platform for next-generation embedded 
non-volatile memory and in-memory computing.

\begin{figure*}
    \centering
        \includegraphics[width=0.8\textwidth]{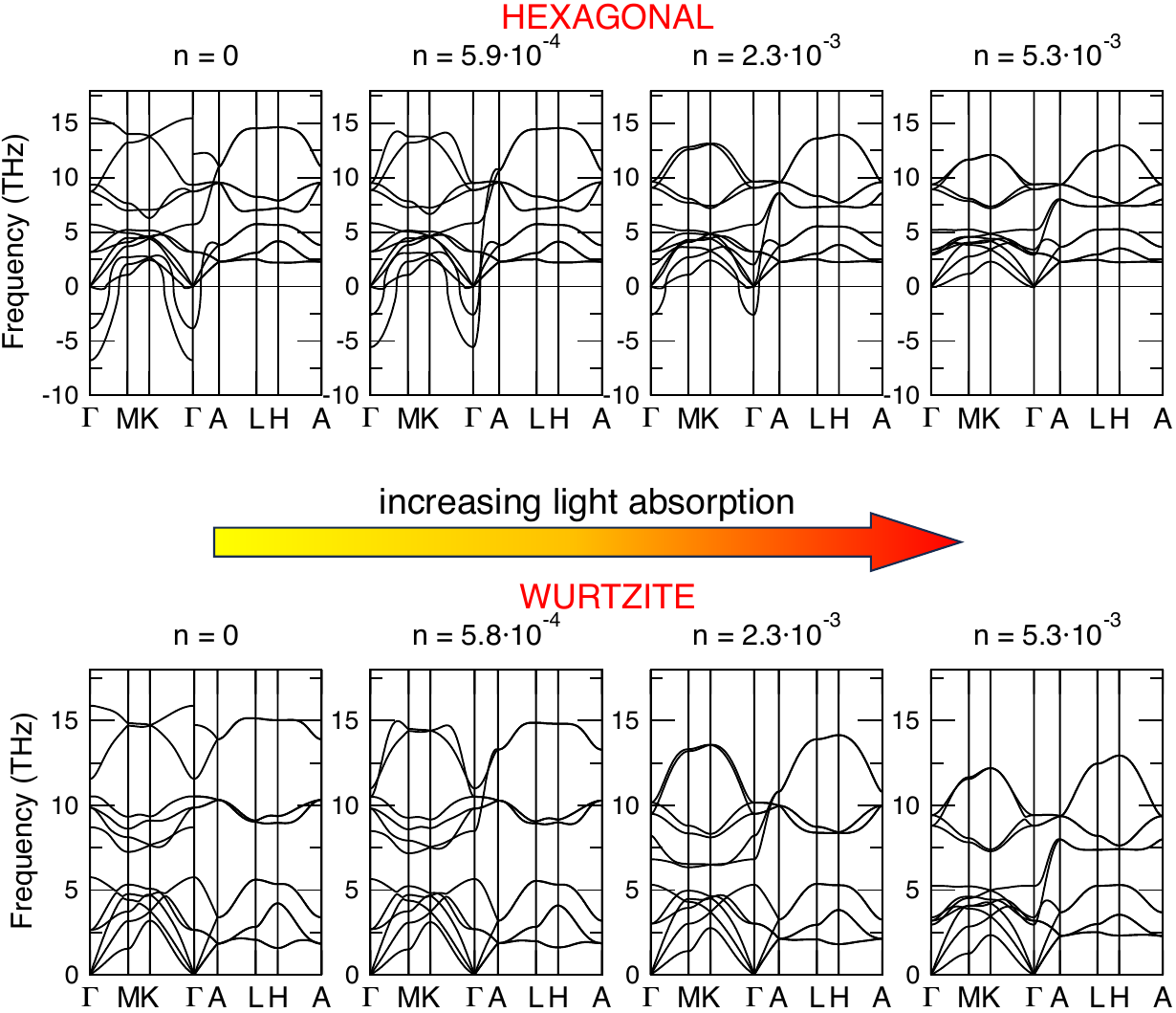}
        \caption{Phonon dispersion of the hexagonal H (top row) and WZ (bottom row) polymorphs under increasing photoexcited charge
        density, $n$, expressed in units of $e$/\AA$^3$.}
        \label{fig2}
\end{figure*}

Despite this promise, WZ ferroelectrics suffer from a key limitation: an intrinsically high coercive field, $\mathcal{E}_{c}$. 
Polarization reversal in wurtzites requires the crystal to traverse a high-energy, nonpolar intermediate state associated with a 
transient hexagonal-layered phase (H, space group $P6_{3}/mc$), in which the nitrogen sublattice adopts an energetically unfavourable 
coordination with neighbouring metal atoms \cite{intro7,rowberg21}. This switching pathway yields an $\mathcal{E}_{c}$ approximately 
an order of magnitude higher than that of conventional oxide ferroelectrics. While a high $\mathcal{E}_{c}$ provides a large switching 
barrier that enhances data retention and suppresses unintended polarization reversal, it simultaneously imposes substantial 
operating-voltage requirements that complicate low-voltage CMOS integration \cite{intro8}. Reducing film thickness alone is not 
a viable workaround, since it compromises the achievable polarization. A central challenge in WZ ferroelectrics is therefore to 
identify effective strategies for lowering the energy barrier to polarization switching, thereby reducing $\mathcal{E}_{c}$.

In this Letter, using first-principles calculations, we propose an original strategy for lowering the energy barrier to
polarization switching in wurtzites: inducing above-bandgap electronic excitations with light. Taking LaN as a representative
wurtzite material, we show that optical absorption considerably reduces the energy difference between the ground-state polar
WZ and the transient nonpolar H phase, even stabilising the latter polymorph under sufficient light irradiation. This behaviour 
is rationalised in terms of partial metallization of the polar WZ crystal, which screens dipole-dipole interactions and thereby 
favours the nonpolar hexagonal phase. Assessing the impact of light absorption on other competing polymorphs, such as the nonpolar 
rocksalt phase (RS, cubic $Fm\bar{3}m$), further corroborates the robustness of the proposed ferroelectric-switching strategy.

First-principles calculations based on density functional theory (DFT) were carried out with the \verb!VASP! code
\cite{KressePRB93,KressePRB99}, using an energy cutoff of $500$~eV, the projector augmented wave method \cite{BlochlPRB94}, and
the generalized gradient approximation \cite{PerdewPRL96} for the exchange-correlation energy. We considered three candidate
crystal structures: wurtzite (WZ, space group $P6_{3}/mmc$), rocksalt (RS, $Fm\bar{3}m$), and layered hexagonal (H, $P6_{3}/mc$)
with AA$'$ stacking (Fig.~\ref{fig1}). The Brillouin zones of the WZ and H phases were sampled with an $8 \times 8 \times 6$
{\bf k}-point grid, while a $10 \times 10 \times 10$ grid was used for the RS phase. Van der Waals dispersion corrections,
necessary to properly describe the H phase, were accounted for with the DFT-D3 method \cite{GrimmeJCP10}. Optoelectronic 
properties were calculated with the range-separated hybrid HSEsol functional \cite{hsesol}. Electric polarizations were estimated 
with the Born effective charges method \cite{born1,born2}.  

\begin{figure}
    \centering
        \includegraphics[width=0.45\textwidth]{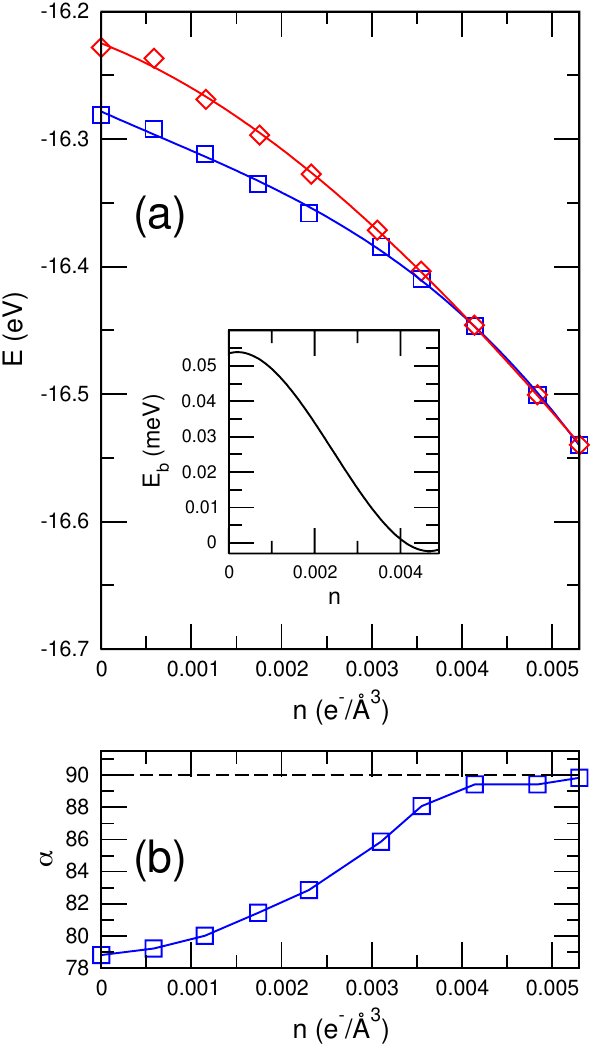}
        \caption{(a)~Energy of the WZ and H polymorphs as a function of photoexcited charge density; the energy difference $E_{b}
        \equiv E_{\rm H} - E_{\rm WZ}$ is shown in the inset. (b)~Evolution of the hexagonal bond angle $\alpha$ in the WZ phase
         under increasing photoexcited charge density; the H phase is characterised by $\alpha = 90^{\circ}$.}
        \label{fig3}
\end{figure}

Photoexcitation was simulated by adjusting the smearing of the Fermi-Dirac distribution, such that the conduction band is 
effectively populated at the expense of depleting the valence band. This effective approach has been successfully employed in a 
series of previous works \cite{CazorlaNanoscale24,RuraliPRL24,RayaMorenonpj2D24,CazorlaAdvFunctMat25}, showing excellent agreement 
with more sophisticated photoexcitation methods \cite{PaillardPRL19}. Phonon dispersions were computed by finite differences with the
\verb!Phonopy! software \cite{TogoJPCM23}, using $4 \times 4 \times 3$ supercells for the WZ and H phases and a $4 \times 4 \times 4$ 
supercell for the RS phase. Finite-temperature renormalized phonons were obtained with the \verb!Dynaphopy! software \cite{CarrerasCPC17} 
by analysing {\it ab initio} molecular dynamics simulations performed with a timestep of $1.5$~fs and total simulation times of 
$\sim 50$~ps. Free energies were calculated within the quasiharmonic approximation \cite{born1,RuraliPRL24}.

\begin{table*}[t]
    \centering
    \begin{tabular}{cccccc}
    \hline
    \hline
  \quad  Material \quad & \quad $e_{33}$ \quad & \quad $E_g$ \quad & \quad $P_{s}$ \quad & \quad $\epsilon$ \quad & \quad $E_{b}$ \quad \\ 
  \quad    (WZ)   \quad & \quad (C~$\cdot$~m$^{-2}$) \quad & \quad (eV) \quad & \quad (C~$\cdot$~m$^{-2}$) \quad & \quad 
  ($\epsilon_{0}$) \quad & \quad (meV/f.u.) \quad \\
    \hline
	    AlN  & 1.64 & 6.1 & 1.20 & 8.6  &  182 \\
	    GaN  & 0.61 & 3.3 & 1.24 & 10.4 &  468 \\
	    LaN  & 2.00 & 2.3 & 0.61 & 17.3 &   58 \\
	    ZnO  & 1.20 & 3.3 & 0.93 & 10.3 &   98 \\
    \hline
    \hline
    \end{tabular}
    \caption{Calculated electronic and dielectric properties of archetypal wurtzite materials. $e_{33}$ represents 
	     the piezoelectric stress tensor component along the $c$-axis, $E_{g}$ the optical bandgap, $P_{s}$ the 
             spontaneous electrical polarization along the $c$-axis, $\epsilon$ the static dielectric tensor averaged 
             over its diagonal components, and $E_{b}$ the energy barrier to polarization switching.} 
    \label{table1}
\end{table*}

Table~1 reports the electronic and dielectric properties calculated for several archetypal wurtzite materials using first-principles 
methods. LaN was selected as the WZ test case for this study for several reasons. First, and most importantly, since we focus on 
light-induced effects, the optical bandgap of LaN ($E_{g} = 2.3$~eV) is ideal for radiation absorption in the visible range and, 
consequently, for practical applications. Second, LaN exhibits superior piezoelectric and dielectric properties compared with AlN, 
GaN, and ZnO (Table~1). For instance, the piezoelectric stress tensor component along the $c$-axis, $e_{33}$, is approximately $67\%$ 
larger in LaN than in ZnO, and $22\%$ larger than in AlN. Likewise, the dielectric constant, $\epsilon$, of LaN is larger than that 
of the other considered WZ materials. In contrast, the spontaneous electric polarization of LaN is the smallest, although still 
sizable and suitable for practical applications ($P_{s} = 0.61$~C$\cdot$m$^{-2}$). In view of these properties, we propose LaN as 
the ideal wurtzite for experimental validation of the theoretical findings presented below.

Figure~\ref{fig2} shows the effect of light irradiation on the vibrational phonon spectrum of LaN in the H and WZ polymorphs. Under 
dark conditions ($n = 0$), the WZ phase is fully vibrationally stable, whereas the H phase is not, as evidenced by several imaginary 
phonon frequencies (shown as negative values in the plot) appearing at the center of the Brillouin zone. This result is consistent 
with the H phase being the saddle point along the ferroelectric switching path of the WZ phase \cite{intro7,rowberg21}. Under increasing 
photoexcitation ($n > 0$), however, the vibrational instabilities of the H phase disappear for photoexcited charge densities equal to 
or larger than $0.005$~$e^{-}$/\AA$^{3}$ (Fig.~\ref{fig2}). In contrast, under increasing $n$ all phonon frequencies of the WZ phase 
remain positive and real, indicating full vibrational stability; more quantitatively, some of the high-frequency WZ optical branches 
($> 10$~THz) shift appreciably toward lower energies, while some of the low-frequency optical branches ($< 5$~THz) shift toward higher 
energies. Interestingly, the phonon spectra of the H and WZ polymorphs turn out to be practically identical at the highest analyzed 
$n$, suggesting collapse of the two structures into a single phase.

The phonon results enclosed in Fig.~\ref{fig2} hint at large structural variations in the considered LaN polymorphs induced by light, 
which in turn may alter the energy barrier to polarization switching in the WZ phase. According to a single-domain free-energy 
Landau-Devonshire model of the form $F(P) = \frac{1}{2} \alpha P^{2} + \frac{1}{4} \beta P^{4} - \mathcal{E} \cdot P$, where $P$ 
represents the electric polarization, $\mathcal{E}$ an external electric field, and $\alpha < 0$ and $\beta > 0$ are parameters, the 
\textit{intrinsic} coercive field necessary to switch the electrical polarization amounts to $\mathcal{E}_{c} = \frac{8}{3\sqrt{3}} 
\cdot (E_b/P_{s})$, with $P_{s}$ being the spontaneous electric polarization and $E_{b} \equiv E_{\rm H} - E_{\rm WZ}$ (Table~1) 
\cite{landau}. Although this estimation typically provides an upper bound for the coercive field observed experimentally (where 
ferroelectric switching proceeds through the nucleation and propagation of domains, which requires smaller fields), it illustrates 
that the energy barrier $E_{b}$, readily accessible via first-principles methods, is an appropriate descriptor for tracking possible 
$\mathcal{E}_{c}$ variations induced by external stimuli.

\begin{figure}
    \centering
        \includegraphics[width=0.485\textwidth]{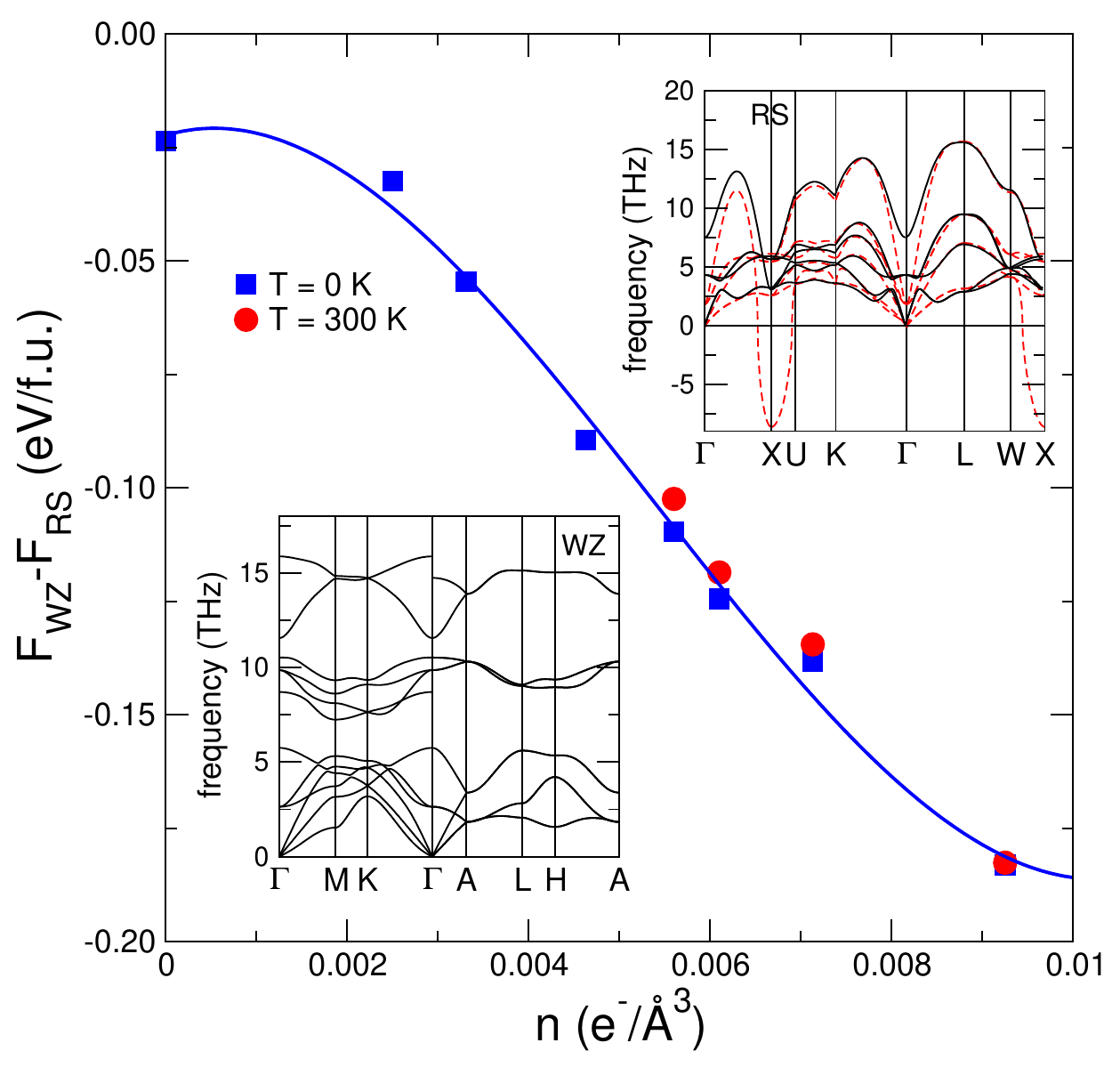}
        \caption{Free-energy difference between the WZ and RS polymorphs, $F_{\rm WZ}$ and $F_{\rm RS}$, at ambient and zero temperature 
         expressed as a function of $n$. Phonon dispersions of the WZ and RS phases are also shown; for the RS polymorph, zero-kelvin 
         (dashed red lines) and room-temperature renormalized phonons (solid black lines) are displayed.}
        \label{fig4}
\end{figure}

Table~1 reports the $E_{b}$ values estimated for different archetypal wurtzite materials under dark conditions ($n = 0$). Among them, 
LaN presents the lowest energy barrier ($58$~meV/f.u.), approximately three times smaller than that of AlN and half that of ZnO. 
Figure~\ref{fig3}a shows the evolution of $E_{b}$ under increasing photoexcited charge density in LaN. The energy difference between 
the two polymorphs is observed to decrease monotonically with increasing $n$, nominally vanishing at values equal to or larger than 
$0.005$~$e^{-}$/\AA$^{3}$. This behavior is consistent with that observed in Fig.~\ref{fig2} for the vibrational phonon spectrum, which 
suggests a collapse of the two structures into a single phase under sufficient light irradiation. These findings demonstrate that 
above-bandgap electronic photoexcitation is an effective strategy for lowering the energy barrier $E_{b}$, and consequently for 
facilitating ferroelectric switching, in wurtzite crystals.

Figure~\ref{fig3}b reports the evolution of the hexagonal bond angle $\alpha$ (Figs.~\ref{fig1}d,e) under increasing photoexcitation. 
Under dark conditions, this angle amounts to $78.7^{\circ}$ in the WZ phase and to $90^{\circ}$ in the H phase. It is observed that 
$\alpha$ steadily increases from the characteristic WZ value at $n = 0$ to the characteristic H value at the maximum considered 
photoexcited density. This result, taken together with those shown in Figs.~\ref{fig2} and \ref{fig3}a, shows that the WZ phase 
spontaneously transforms into the H polymorph as a result of increasing above-bandgap electronic photoexcitation, thus significantly 
depleting $E_{b}$. This behavior, in analogy to that previously observed for polar oxides \cite{CazorlaNanoscale24,RuraliPRL24,
RayaMorenonpj2D24,CazorlaAdvFunctMat25,PaillardPRL19}, can be explained in terms of the screening of dipole-dipole interactions in 
polar phases caused by partial light-induced metallization, which energetically favors the related nonpolar phases.

Thus far in this study, we have focused on the impact of light absorption on the phase competition between the WZ and H polymorphs 
of LaN, the two phases directly involved in ferroelectric switching. However, other phases, such as the cubic nonpolar rocksalt 
structure (RS, $Fm\bar{3}m$), can also be thermodynamically and compositionally stabilized in binary nitrides and oxides under 
ordinary dark conditions \cite{rocksalt}. It is therefore legitimate to ask whether light absorption similarly affects these 
competing phases, and whether such effects might alter the conclusions presented above.

Figure~\ref{fig4} shows the free-energy difference between the WZ and RS phases expressed as a function of photoexcited charge 
density and temperature. Finite-temperature effects were taken into account with the quasiharmonic approximation, considering 
$T$-renormalized phonons \cite{born1,RuraliPRL24}. Under zero-temperature conditions, the energy of the ground-state WZ phase 
is approximately $20$~meV per formula unit smaller than that of the RS phase. Under increasing photoexcitation, this energy 
difference increases in absolute value, with the WZ polymorph remaining the ground state. Likewise, when thermal effects are 
considered at room temperature, the free-energy difference $F_{\rm WZ} - F_{\rm RS}$ follows the same trend as the corresponding 
internal energy difference, both qualitatively and quantitatively. These results indicate that light-induced effects on WZ 
ferroelectric switching are largely unaffected by the evolution of other competing nonpolar polymorphs under $n \neq 0$ conditions.

In summary, using first-principles calculations we have demonstrated that above-bandgap electronic photoexcitation offers an 
original and effective route for lowering the energy barrier to ferroelectric switching in wurtzite crystals, taking LaN as a 
representative test case (analogous results, not shown here, have also been obtained for ZnO). Light-induced carriers stabilize 
the nonpolar hexagonal phase that mediates polarization reversal relative to the wurtzite ground state, thereby significantly 
reducing $E_{b}$. This behavior, which we attribute to a photoinduced partial metallization that screens the dipole-dipole 
interactions stabilizing ferroelectric order, is unaffected by the light-induced evolution of other competing nonpolar phases, 
underscoring the robustness of the mechanism. Because the intrinsic coercive field scales directly with this energy barrier, our 
results indicate that optical excitation can substantially reduce the operating voltages currently required for WZ-based 
ferroelectric switching, without the drawbacks associated with reducing film thickness. Beyond LaN, given the broad chemical 
tunability of the wurtzite family, we expect this light-facilitated switching mechanism to be readily transferable to other 
technologically relevant nitrides and oxides. These findings establish light as a non-invasive, low-power complement to conventional 
electric-field control, opening new avenues for faster, lower-voltage, and more energy-efficient non-volatile memory and neuromorphic 
technologies based on wurtzite ferroelectrics.

\section*{Acknowledgments}
R.R. acknowledges support by MICIN/AEI/10.13039/501100011033 under grant PID2024-162811NB-I00, and the Severo Ochoa Centres of 
Excellence Program under grant CEX2019-000917-S. C.C. acknowledges support by MICIN/AEI/10.13039/501100011033 and ERDF/EU under 
the grants CNS2025-165467, PID2023-146623NB-I00, PID2023-147469NB-C21 and the Maria de Maeztu Units of Excellence Programme 
CEX2023-001300-M, and by the Generalitat de Catalunya under the grants 2021SGR-00343, 2021SGR-01519 and 2021SGR-01411. Computational 
support was provided by the Red Española de Supercomputación under the grants FI-2024-1-0005, FI-2024-2-0003, FI-2024-3-0004, 
FI-2024-1-0025, FI-2024-2-0006, FI-2024-1-0012, FI-2024-1-0015 and FI-2025-2-0015, and the Centro de Supercomputaci\'on de Galicia 
(CESGA).
\\

\bibliography{references} 

\end{document}